\documentclass[reprint,
 amsmath,amssymb,
 aps,
pra,
]{revtex4-2}
\usepackage{tabularx}
\usepackage{threeparttable}
\usepackage{graphicx}%
\usepackage{dcolumn}%
\usepackage{bm}%
\usepackage{braket}

\usepackage{hyperref}
\hypersetup{
    colorlinks=true,
    linkcolor=blue,
    urlcolor=blue,
    citecolor=blue
}
\usepackage{xcolor}
\usepackage{soul}
\usepackage[normalem]{ulem}

\begin{document}
\preprint{APS/123-QED}

\title{Tantalum-Encapsulated Niobium Superconducting Resonators: High Internal Quality Factor and Improved Temporal Stability via Surface Passivation}

\author{Anas Alkhazaleh}
\email{Anas.Alkhazaleh@tii.ae}
\author{Juan Villegas}
\author{Florent Ravaux}
\author{Alexey Zharinov}

\affiliation{%
 Quantum Research Centre, Technology Innovation Institute\\ 
 Abu Dhabi, United Arab Emirates
}%

\date{\today}

\begin{abstract}

Superconducting coplanar waveguide resonators are essential components in quantum processors, where their internal quality factor (\(Q_{i}\)) constrains qubit coherence and readout fidelity. In niobium devices, microwave losses at millikelvin temperatures are strongly influenced by two-level systems (TLS) associated with the complex NbO\(_x\) surface oxide. To mitigate these losses, we investigate a surface-engineering approach in which Nb films are capped \textit{in situ} with a thin tantalum layer to suppress Nb\(_2\)O\(_5\) formation and replace the native NbO\(_x\) interface with a Ta-based oxide.

We fabricate Nb/Ta bilayer and reference Nb resonators on high-resistivity silicon using identical DC sputtering and wet etching conditions, and characterize their performance at millikelvin temperatures. Fresh Ta-encapsulated devices exhibit internal quality factors up to \(2.4 \times 10^{6}\) in the near--single-photon regime, with power dependence consistent with reduced TLS-related loss at the metal--air interface. A control Nb device fabricated under the same process shows comparatively lower \(Q_{\mathrm{TLS}}\), consitent with the beneficial effect of the Ta capping layer. Furthermore, ageing tests performed on Nb/Ta resonators after six months reveal a moderate reduction in \(Q_{\mathrm{TLS}}\) relative to their initial values, yet the performance remains superior to newly fabricated Nb-only devices. These results suggest that thin Ta encapsulation enhances interface quality and contributes to improved temporal stability while remaining compatible with Nb-based fabrication workflows.

\end{abstract}

\maketitle

\section{Introduction}
Superconducting coplanar waveguide (CPW) resonators serve as critical components in quantum information processing, functioning as readout elements for superconducting qubits, microwave-photon storage and emitter, and as testbeds for characterizing material loss mechanisms at microwave frequencies~\cite{McRae_2020,Zmuidzinas_2012,Gao_2008_Thesis}. Over the past few years, material and process optimization have helped to push quality and coherence times of quantum superconducting circuits from tens of microseconds into the hundreds-of-microseconds to millisecond regimes, using emerging materials like tantalum~\cite{Place_2021,Ganjam_2024,Bland_2025,Abdisatarov_2025}. The performance of these resonators, quantified by their internal quality factor $(Q_{i})$, directly impacts qubit coherence times and measurement fidelity in quantum processors.

Microwave loss in superconducting CPW resonators at millikelvin temperatures and in the single-photon regime is typically dominated by two-level systems (TLS) in amorphous interfacial dielectrics, with additional contributions from non‑TLS mechanisms depending on device design and fabrication. In contrast, resonators based on alternative high-impedance materials, such as granular aluminum (GrAl), can exhibit dissipation dominated by non-equilibrium quasiparticles rather than TLS, with reported values of $Q_i \sim 10^5$ in the single-photon regime~\cite{Grunhaupt_2018}. 

For planar CPWs, experiments indicate that the largest participation comes from the metal--substrate (MS) interface (including native metal oxides and amorphous interlayers on substrate), metal--air (MA) interface (comprising native oxides and surface adsorbates) and substrate--air (SA) interface, especially native SiO$_2$ or damaged surface regions~\cite{Marcaud_2025,Premkumar_2021,Pappas_2011,Gao_2008,deGraaf_2020,Altoe_2022}. Here, the TLS loss mechanism can be modeled as dipoles that couple to the resonator electric field and contribute a loss tangent with characteristic power and temperature dependence \cite{Martinis_2005,Gao_2008,Niepce_2019,Faoro_2012}.

While niobium has been a widely used material for superconducting quantum circuits due to its relatively high critical temperature (\(T_{c} \approx 9.2\) K), large bandgap (1.38 meV at 0 K), mechanical robustness, and well-established fabrication techniques, its native surface oxide, primarily Nb$_2$O$_5$, has emerged as a dominant source of microwave dissipation, limiting resonator performance in the quantum regime. 

Detailed structural and spectroscopic investigations show that Nb exposed to typical processing and ambient conditions develops a multilayered, mixed-valence oxide (NbO, NbO$_2$, Nb$_2$O$_5$) with significant structural disorder and a graded interface into the metal. These oxides are structurally complex, chemically active, hydride-forming, and often highly lossy at microwave frequencies. This oxygen-rich region supports TLS and can trap hydrogen and other impurities~\cite{Oh_2024,Torres-Castanedo_2024,Wang_2024}. Early systematic studies on Nb and Al CPWs showed that the increased loss at low power and low temperature can be explained by TLS located at surfaces and interfaces rather than in the bulk substrate~\cite{Pappas_2011,Gao_2008}. 

By varying the CPW gap width \(g\), center-conductor width \(s\), and trench geometry, the authors show that interface participation ratios strongly affect loss; in their Nb-on-Si resonators, the MS interface has the largest loss participation, while selective etching of the substrate-air and metal-air oxides reduces both TLS and non-TLS losses \cite{Altoe_2022}. Even when the base film is high-quality and epitaxial, Wang and collaborators~\cite{Wang_2024} showed that variations in HF-based oxide removal and acid cleaning protocols can significantly modify the near-surface microstructure and induce large changes in $Q_i$ and residual resistivity. This is why fabrication steps that change only the surface chemistry, such as different etchants, HF dips, or surface passivation, can strongly alter resonator loss without appreciably affecting the bulk~\cite{Pappas_2011,Wang_2024,Lozano_2024,Faoro_2012,Muller_2019}.

These observations are consistent with broader materials reviews: the loss in Nb devices is often limited by the details of the oxide and hydride structure near the surface, rather than the intrinsic superconducting properties of the bulk~\cite{Altoe_2022,Premkumar_2021}. It is therefore natural to consider strategies that either remove or bypass the Nb surface oxide, or reduce its participation in the resonator fields.

Tantalum is increasingly viewed as a particularly promising material for superconducting quantum circuits. Structural and spectroscopic characterizations provide a rationale for these device-level improvements, showing that Ta tends to form a more uniform, stable Ta$_2$O$_5$-based surface oxide with a sharper metal--oxide interface and fewer mixed-valence suboxides than Nb~\cite{Oh_2024,McLellan_2023}. The superior performance of Ta$_2$O$_5$ compared to Nb$_2$O$_5$ as a surface oxide is attributed to several factors. First, Ta$_2$O$_5$ forms with more stoichiometric composition and reduced structural disorder compared to the variable oxygen content observed in Nb oxides~\cite{Oh_2024,McLellan_2023,Bal_2024,Premkumar_2021}. The predominantly stoichiometric Ta$_2$O$_5$ layer is expected to reduce the potential for magnetic moment formation and TLS‑related loss compared to the non-stoichiometric niobium oxide~\cite{Faoro_2012,deGraaf_2020,Premkumar_2021,Bal_2024}. Comparative structural analysis suggests that the amorphous Ta$_2$O$_5$ phase, while still disordered, exhibits improved chemical and structural uniformity relative to Nb$_2$O$_5$~\cite{Oh_2024,McLellan_2023,Bal_2024}. Experiments using Ta as base layer showed that replacing Nb with Ta yields high coherence times, close to milliseconds~\cite{Place_2021,Ganjam_2024,Bland_2025}, and very high Q\textsubscript{i} values of CPW above (10\textsuperscript{6}) ~\cite{Marcaud_2025,Lozano_2024,Urade_2024,Poorgholam-Khanjari_2025}. Overall, these studies support the view that $\alpha$-Ta, with a well-formed Ta$_2$O$_5$ surface, is intrinsically less lossy than Nb with its complex NbO$_x$ under comparable processing conditions, and that Ta's loss mechanisms can be more controllable.

Surface encapsulation has emerged as a good alternative to leverage the benefits of using Nb while damping the major loss coming from the surface oxides of the niobium where another superconducting material is used to passivate Nb before NbO$_x$ forms~\cite{Bal_2024}. In recent work~\cite{Bal_2024} an encapsulation strategy was developed in which Nb films are capped in situ with a thin Ta, Al, TiN, or Au film before exposure to ambient. Accross different fabrication lines, substrates, and multiple capping materials, they found that encapsulated Nb transmons exhibit relaxation time (T$_1$) values 2 to 5$\times$ longer than baseline devices, with Ta caps offering median T$_1$ above 300 $\mu$s and up to around 600 $\mu$s~\cite{Bal_2024}. Because the only change is the surface structure (the underlying Nb, geometry, and junction technology are held fixed), this work isolates the role of the surface oxide and strongly implicates NbO$_x$ as a high-loss dielectric compared with Ta$_2$O$_5$, Al$_2$O$_3$, or TiN$_x$O$_y$ ~\cite{Bal_2024}.

These results motivate hybrid material approaches in which tantalum is used as a surface-engineering layer rather than as a bulk material, especially on silicon wafers where directly maintaining the $\alpha$ phase of Ta is not straightforward. In particular, bilayer stacks composed of a thick Nb base layer and a thin Ta cap leverage the mechanical robustness, process maturity, and existing fabrication infrastructure optimized for Nb, while exposing the microwave fields primarily to a Ta-terminated surface. Recent work on Ta resonators deposited on Nb seed or buffer layers demonstrates that composite Nb/Ta thin films can support $\alpha$-Ta growth and achieve high $Q_i$ values without substrate heating~\cite{Urade_2024,Poorgholam-Khanjari_2025}. 

Building on this, the present work investigates how Ta encapsulation affects resonator quality factors across different power regimes, with a focus on quantum-limited readout. The central hypothesis is that \textit{in situ} Ta encapsulation of Nb coplanar waveguide resonators strongly suppresses the formation of lossy Nb$_2$O$_5$ at the metal--air interface, thereby reducing TLS-mediated losses and enhancing internal quality factors in the quantum regime. In this picture, a 25--35~nm Ta capping layer deposited without breaking vacuum acts as a diffusion and oxidation barrier for the underlying 165--175~nm Nb film, substantially reducing the Nb$_2$O$_5$ that would otherwise form at the exposed surface and replacing it with a Ta$_2$O$_5$-terminated interface. Because Ta$_2$O$_5$ hosts fewer and less deleterious TLS than the complex, mixed-valence NbO$_x$ stack, the dominant TLS loss channel at the metal--air boundary is expected to be suppressed, which in turn is anticipated to improve $Q_i$ for Ta/Nb bilayer resonators, particularly in the low-photon-number regime~\cite{Dhundhwal_2025,Arabi_2024,Krasnikova_2025}.

\section{Resonator Design and Electromagnetic Modeling}

For the present devices, the external coupling rate $\kappa_c$ is designed to exceed the internal loss rate, placing the resonators in a strongly overcoupled regime rather than at critical coupling ($Q_e \ll Q_i$, where $Q_e$ represents the external quality factor). This choice ensures pronounced extinction ratios with deep transmission dips across all measured resonators, enabling robust fitting of the resonance lineshape and reliable extraction of $Q_i$. Although the loaded quality factor is set primarily by the external coupling in this regime, the large separation between $Q_e$ and $Q_i$ still allows the intrinsic dissipation to be determined with good  
accuracy. As a result, the extracted $Q_i$ remains a meaningful probe of microscopic energy--loss mechanisms, including contributions from two--level systems, under otherwise identical measurement conditions.

The loaded quality factor $Q_l$ of a resonator is determined by the stored energy $E$ relative to the energy dissipated per cycle $P_{diss}$,
\begin{equation}
    Q_l = \frac{\omega_r E}{P_{diss}} = \left( \frac{1}{Q_e} + \frac{1}{Q_i} \right)^{-1}.
    \label{eq:Q_load}
\end{equation} 
In the absence of coupling to any additional modes, this model offers a complete representation of the resonator dynamics~\cite{Gao_2008_Thesis,Goppl_2008}. The external quality factor $Q_e$ is engineered by the coupling network to set the resonator decay rate, qubit Purcell relaxation, and readout signal-to-noise ratio, whereas the internal quality factor $Q_i$ quantifies intrinsic dissipation arising from multiple microscopic channels~\cite{Gao_2008_Thesis,Goppl_2008,Zmuidzinas_2012,Khalil_2012,McRae_2020}. Prominent contributions include dielectric loss from two-level systems $\delta_{\text{TLS}}$~\cite{Martinis_2005,Gao_2008,Pappas_2011,Faoro_2012}, loss due to nonequilibrium quasiparticles $\delta_{\text{QP}}$(including subgap trapping)~\cite{deGraaf_2020,Ganjam_2024,Fischer_2023,Fischer_2024,Fischer_2025}, radiative or packaging-related loss $\delta_{\text{rad}}$~\cite{Huang_2021,Lienhard_2019,Bruno_2015}, and losses associated with magnetic vortices $\delta_{\text{vortex}}$~\cite{Lindstrom_2009,Altoe_2022,Abdisatarov_2025}, whose relative importance depends on materials, geometry, shielding, and operating conditions~\cite{McRae_2020,Premkumar_2021,Zmuidzinas_2012}.

\begin{figure}[!h]
\includegraphics[width=\columnwidth]{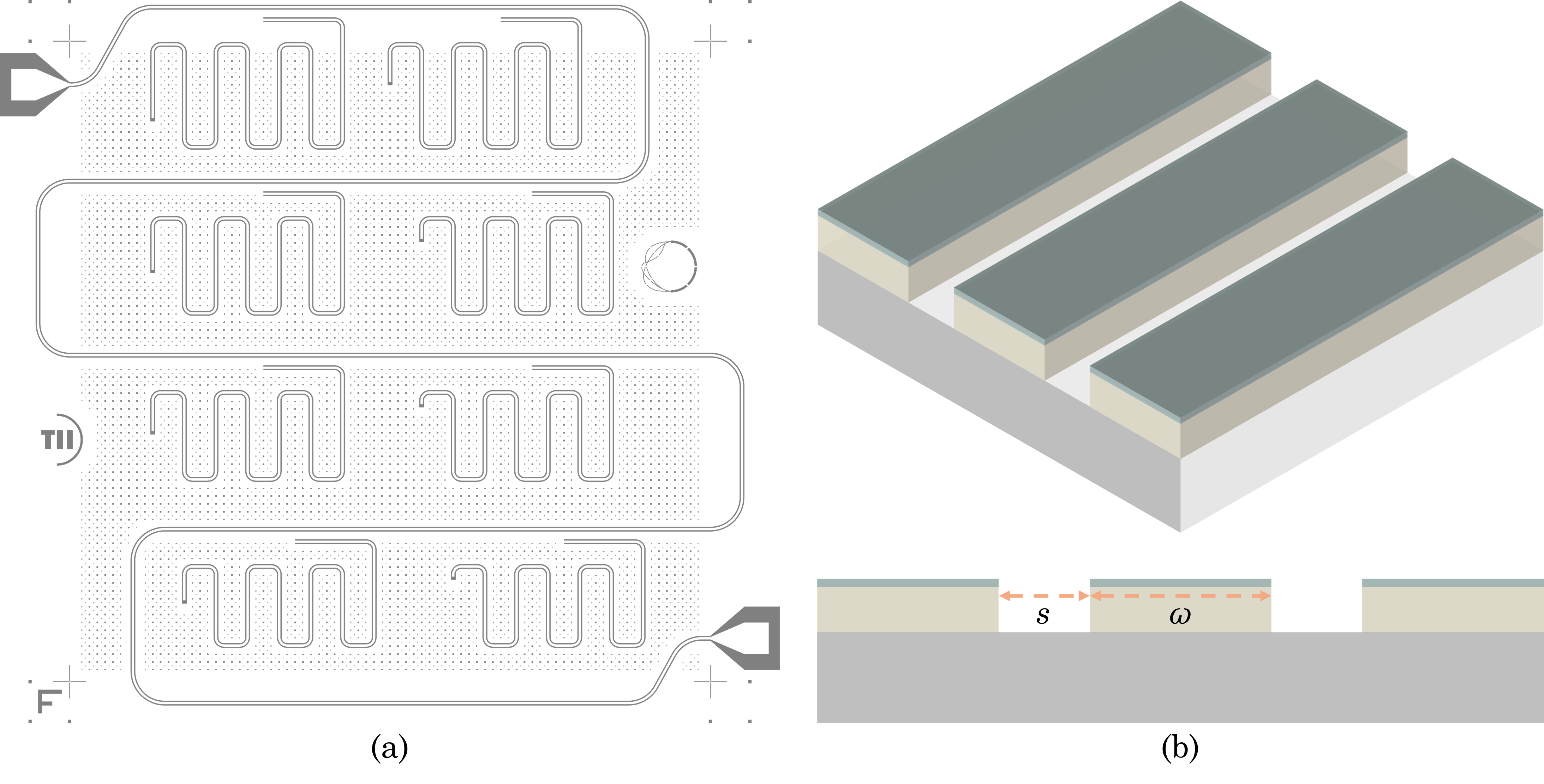}
\caption{(a) Layout of eight $\lambda/4$ resonators and (b) CPW architecture.}
\label{fig:res_layout}
\end{figure}

Here, the resonators are formed in a $\lambda/4$ configuration to support a mode with short overall length by imposing a short-circuit boundary condition in one end, and an open circuit in the other, see Fig.~\ref{fig:res_layout}. They are built in a coplanar waveguide configuration with a central width of $15.0 \mu m$ and a gap of $7.5 \mu m$, which for a silicon substrate generates a waveguide with characteristic impedance of $Z_0 = 47.9\ \Omega$ and relative effective permittivity $\varepsilon_{eff} = 6.34$. The resonance of the $n^{th}$ mode in a waveguide resonator is calculated from \cite{Goppl_2008}
\begin{equation}
    \omega_n = \frac{1}{\sqrt{L_n(C+2C^*)}},
    \label{eq:resonance_freq}
\end{equation}
where $L_n$ and $C$ are  the equivalent inductance and capacitance of the resonator, derived from the waveguide's inductance ($L_{\ell}$) and capacitance ($C_{\ell}$) per unit length such that, $C = C_{\ell} \ell/2$ and $L_n = 2L_{\ell}\ell/(n\pi)^2$, with $\ell$ being the resonator length. The term $C^*$ is the Norton equivalent capacitance arising from the capacitor that connects the resonator to the feedline given by
\begin{equation}
    C^* = \frac{C}{ 1+ \omega_n^2 C_k^2 R_L^2},
    \label{eq:C_norton}
\end{equation}
with $C_k$ the coupling capacitance, and $R_L$ the load (feedline) impedance. Note that, in Eq.~\ref{eq:resonance_freq}, the factor of 2 accounts for the imposed boundary conditions; this is the case for either half or quarter wave resonators. The external quality factor is primarily controlled by this $C_k$ and can be modeled as
\begin{equation}
    Q_e = C \frac{1+ \omega_n^2C_k^2R_L^2}{2\omega_nC_k^2R_L} \approx \frac{C}{2\omega_nC_k^2R_L},
    \label{eq:Q_external}
\end{equation}
since the coupling capacitor reactance is designed to be small relative to the load impedance, providing $\omega_n^2 C_k^2 R_L^2\sim 0.0006$ for the fundamental mode. The values of $C_k$ are extracted from finite element (FEM) electrostatic simulations. The estimated values for $C_k, C,~{\rm and}~ L_n$ are used only to target the resonator performance during design. In the experimental analysis presented in section \ref{sec:setup}, we extract all resonance parameters from the complex transmission coefficient $S_{21}$ spectroscopy fitting.

\section{Devices Fabrication}
\begin{figure*}[!t]
  \centering
  \includegraphics[width=\textwidth]{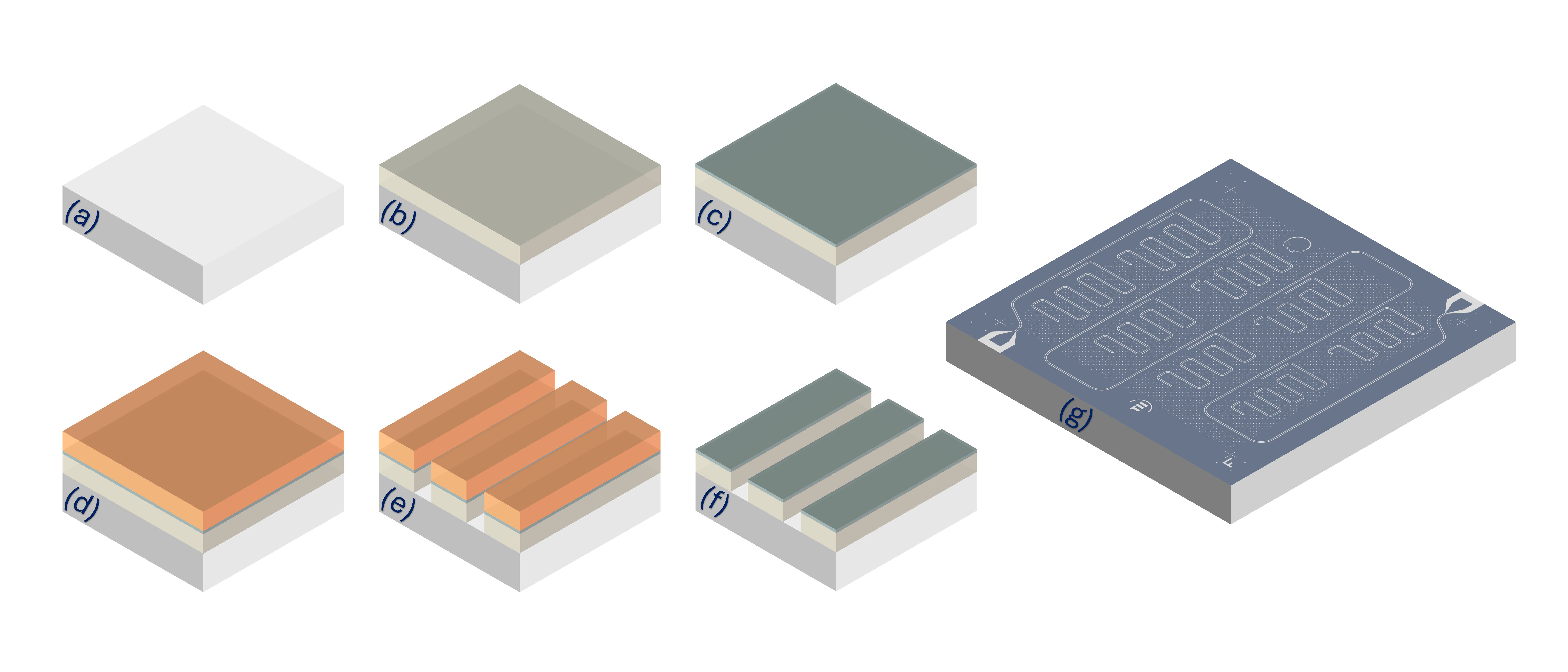}
  \caption{\label{fig:Fab_steps}Fabrication steps of CPW Resonators, (a) substrate cleaning, (b--c) Nb/Ta bilayer sputtering, (d--f) photolithography and wet etching, and (g) diced chips after cleaning.}
\end{figure*}

Devices were fabricated on high-resistivity (\textgreater10 k$\Omega$·cm) Float Zone (FZ) intrinsic Si(100) substrates. Prior to processing, 4-inch wafers were diced into smaller pieces to enhance handling and process reproducibility. Surface preparation began with a 15-minute oxygen plasma descum to remove organic contaminants, followed by a hot N-Methyl-2-pyrrolidone (NMP, 80\,$^\circ$C) bath to remove any organic residues. The substrates were then sonicated sequentially in acetone and isopropyl alcohol (IPA), dehydrated by baking, and cooled under ambient conditions. Immediately before loading into the sputtering tool, the native silicon oxide was removed by dipping in 1\,\% hydrofluoric acid (HF), followed by deionized water (DIW) rinsing and nitrogen (N$_2$) drying.

\begin{figure}[h]
  \centering
  \includegraphics[width=\linewidth]{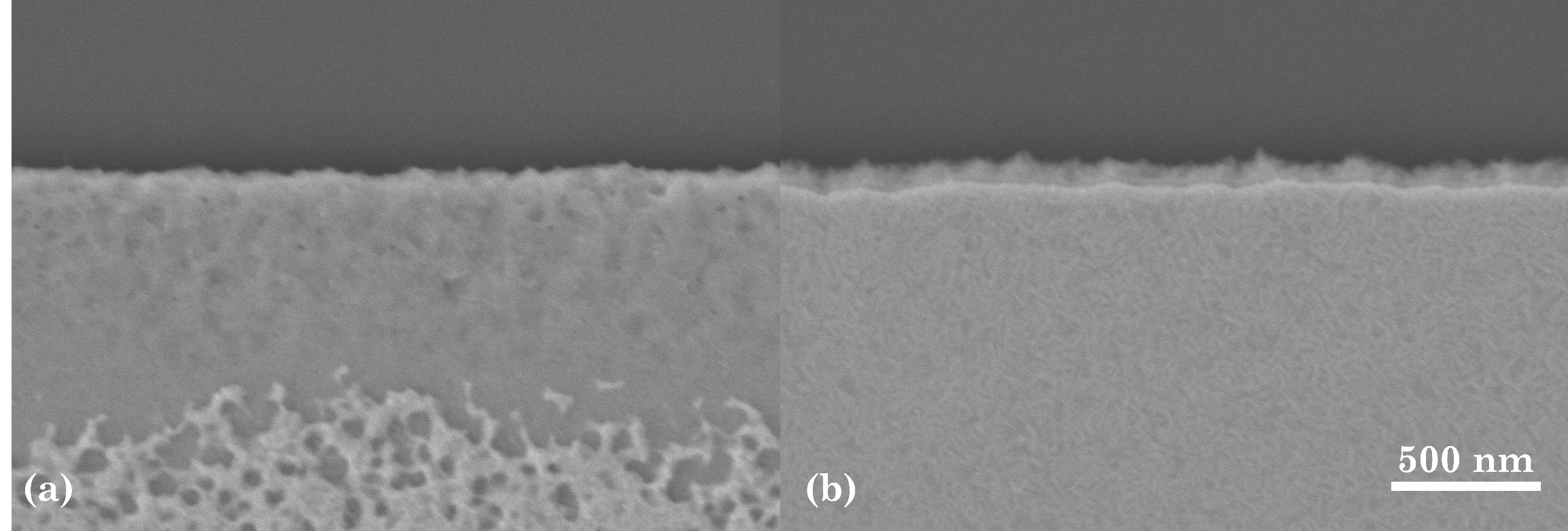}
  \caption{\label{fig:Etch_optm}SEM images showing the effect of etching-parameter optimization on metal bilayer overlap and lateral etching: (a) over-etched Ta accompanied by Nb damage, and (b) reduced lateral etching of the Ta layer following optimization.}
\end{figure}

Metallization was performed in a DC magnetron sputtering system. Samples were transferred from the load-lock to the main chamber once the vacuum reached around $1 \times 10^{-7}$\,mbar. To desorb volatile species, the substrate was baked at 200\,$^\circ$C for 60\,minutes and allowed to cool slowly to room temperature overnight. With a base pressure of approximately $1 \times 10^{-8}$\,mbar, a niobium (Nb) base layer was deposited at a rate of 0.1\,nm/s, followed \textit{in situ} by a tantalum (Ta) capping layer sputtered at 0.2\,nm/s. The total bilayer thickness was approximately 185\,nm. Following deposition, samples were held under vacuum to stabilize before venting the chamber.

Immediately after venting, substrates were spin-coated with a positive photoresist that served both as a passivation layer and a lithography mask. Patterning was performed using a direct laser writer (DLW), with exposure dose and development parameters optimized for high-resolution feature definition across multiple samples. After development, a mild oxygen plasma cleaning removed any residual resist from the exposed regions. A subsequent hard bake enhanced the photoresist's etch resistance.

Pattern transfer was carried out using a wet chemical etch. Both Nb and Ta layers were etched using a single-bath process derived from a diluted Tantalum Etchant\,III solution, calibrated for compatible etch rates of the two metals. After etching, samples were rinsed in sequential DIW baths and dried with N$_2$. The photoresist mask was stripped in a hot NMP bath (1\,hour), followed by acetone and IPA cleaning. Figure \ref{fig:Fab_steps} illustrates the main fabrication steps. The isotropic etch properties were carefully optimized to minimize lateral undercut while ensuring complete removal of exposed bilayer regions, as shown in Fig.~\ref{fig:Etch_optm}.

\begin{figure}[h]
  \centering
  \includegraphics[width=\linewidth]{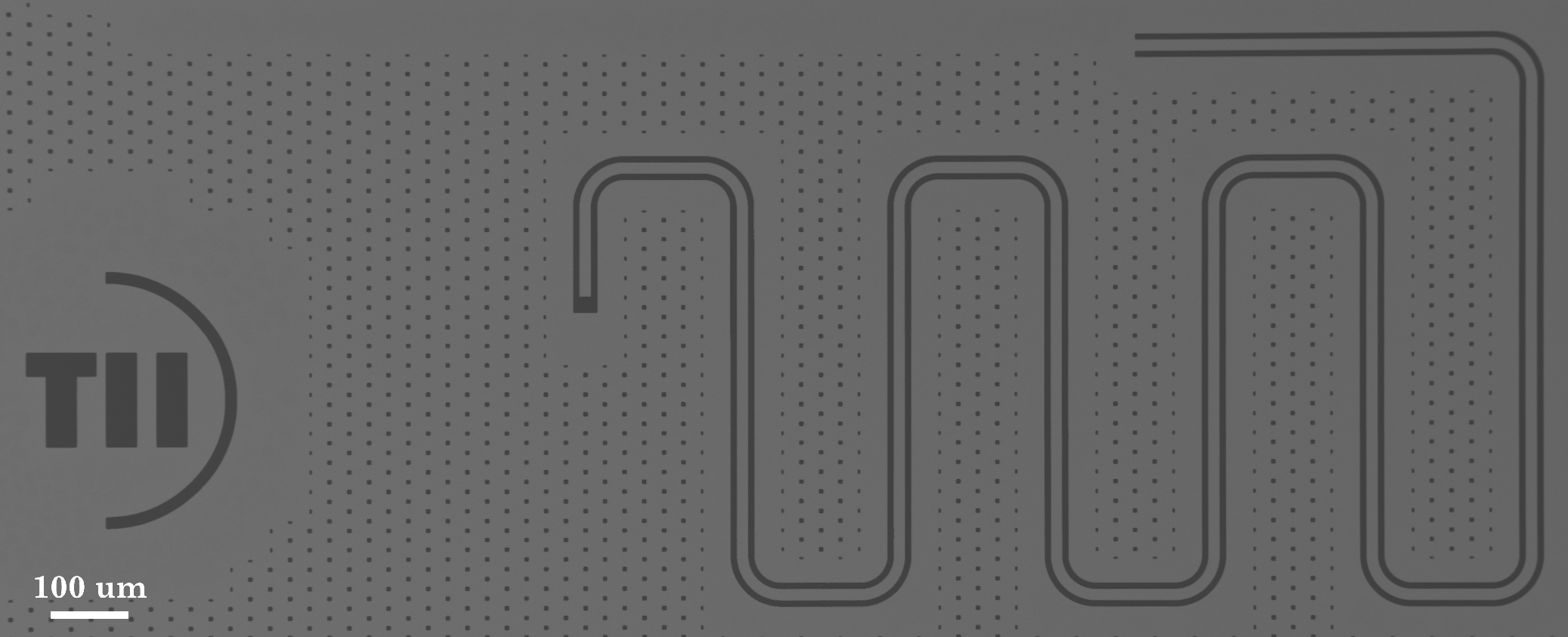}
  \caption{\label{fig:res}SEM Image of quarter wave resonators after etching.}
\end{figure}

\begin{figure}[h]
  \centering
  \includegraphics[width=\linewidth]{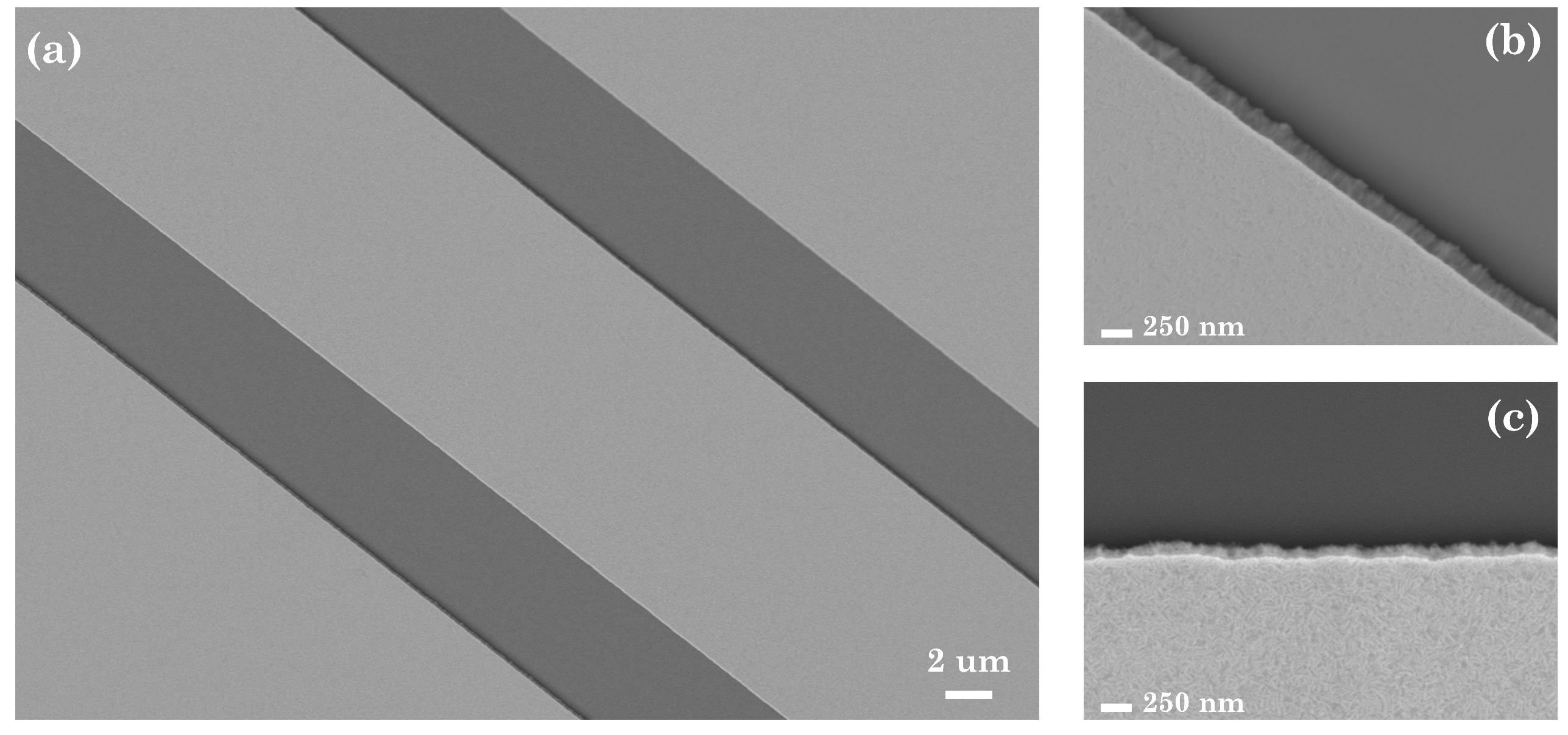}
  \caption{\label{fig:sem_top_angle}Oblique-view SEM image : sidewall profile and degree of lateral etching.}
\end{figure}

Initial process verification was performed using optical microscopy to confirm pattern fidelity. Scanning electron microscopy (SEM) provided detailed inspection of the etched features, enabling assessment of sidewall morphology, metal residues, and critical dimension accuracy, see Fig. \ref{fig:res}. Figure \ref{fig:sem_top_angle} presents an oblique-view SEM image highlighting the sidewall profile and degree of lateral etching.

Prior to dicing, the substrate surface was coated with a protective resist to prevent particle deposition and mechanical damage during dicing. The diced chips were subsequently cleaned in hot NMP, acetone, and IPA baths. Each die was mounted on a Rogers\,TMM10i with bare copper surface PCB, and electrical connections were made using aluminum wedge wire bonding. The assemblies were enclosed in a package consisting of a copper base for thermalization and an aluminum lid for electromagnetic shielding.

\section{Experimental Setup and Measurement Protocol}
\label{sec:setup}

\begin{figure}[h]
  \centering
  \includegraphics[width=\columnwidth]{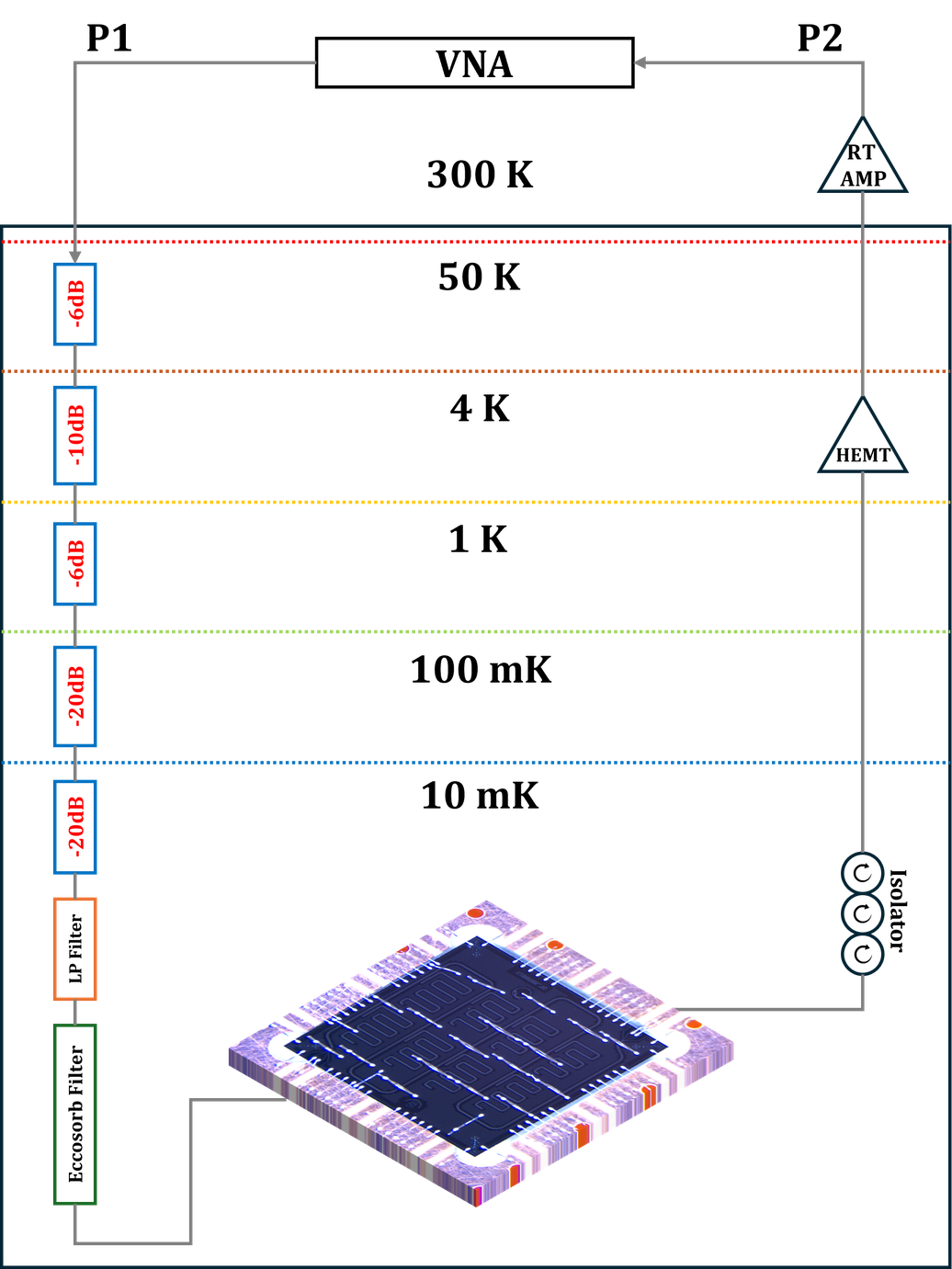}
  \caption{\label{fig:wiring}Wiring scheme of the tested chips.}
\end{figure}

The packaged devices were mounted on the mixing chamber stage of a dilution refrigerator operating at a base temperature of approximately 10\,mK. The microwave input lines incorporated distributed attenuators at various temperature stages to suppress thermal noise, as shown in Fig. \ref{fig:wiring}. Output signals were amplified using a low-noise high-electron-mobility transistor (HEMT) amplifier at the 4\,K stage, followed by additional amplification at room temperature.

Microwave transmission spectroscopy was performed using a Keysight Vector Network Analyzer (VNA). The complex transmission coefficient S$_2$$_1$ was measured to characterize the resonator response. The VNA output power was swept from -30\,dBm to -80\,dBm, and the frequency range was adjusted to cover all resonances of interest. 

\begin{figure}[!h]
  \centering
  \includegraphics[width=\columnwidth]{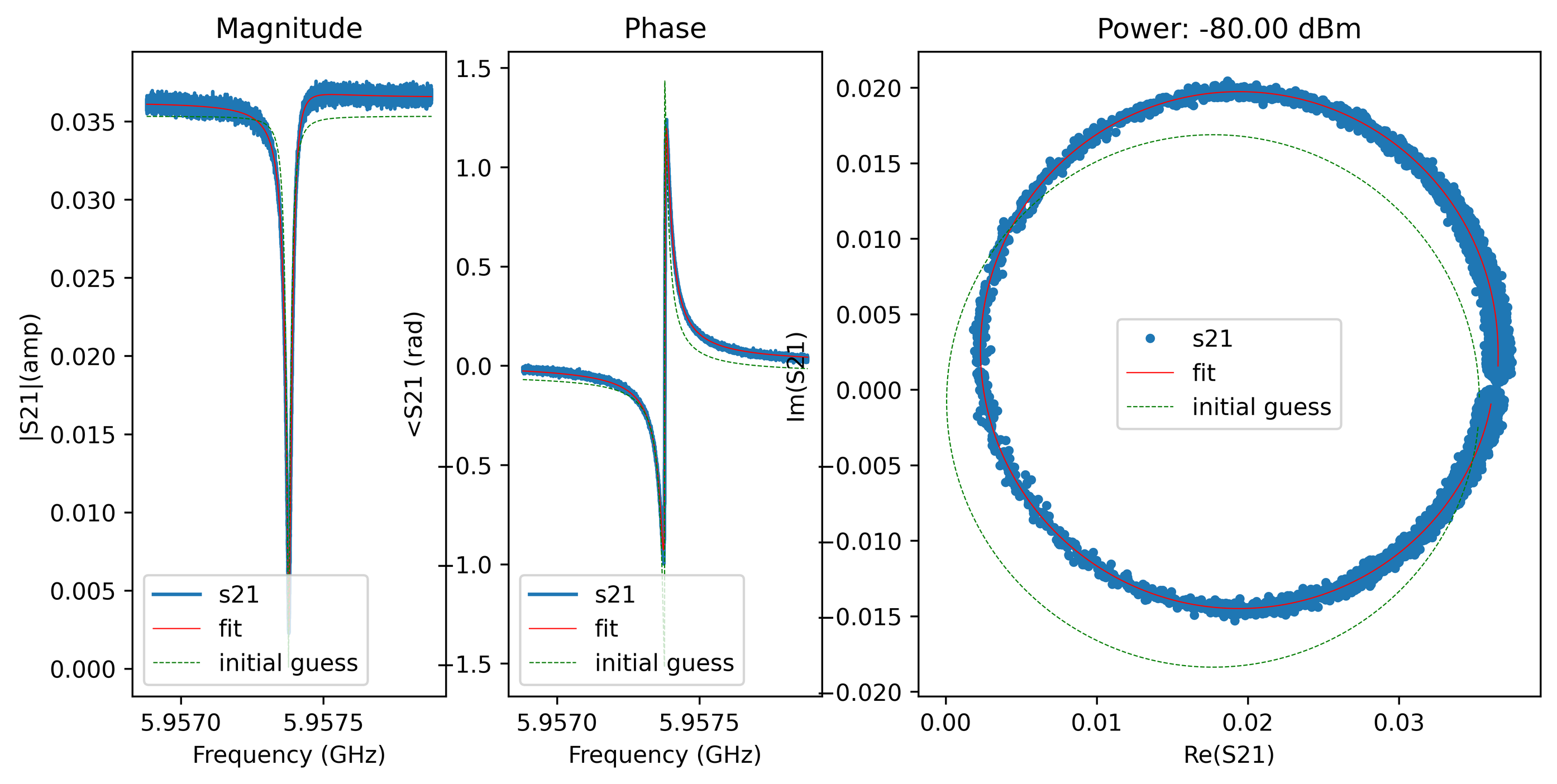} 
  \caption{\label{fig:fitting}S21 fitting of Resonator R5 at -80 dBm [VNA Power].}
\end{figure}

Resonance features were identified as (asymmetric) Lorentzian dips in the transmission magnitude \textbar S$_2$$_1$\textbar{} accompanied by the characteristic phase shift in arg(S$_2$$_1$). The loaded $Q_l$ and external $Q_e$ quality factors were extracted by fitting the complex S$_2$$_1$ data by fitting to 
\begin{equation}
    S_{21}(f) = a(f)  \left[1 - \frac{\frac{Q_l}{|Q_e|} e^{i\theta}}{1 + 2i Q_l \left( \frac{f - f_r}{f_r} \right)} \right],
    \label{eq:S21_model}
\end{equation}
where $a(f)$ is the modified complex transmission amplitude (see S4 in Ref. \cite{Bruno_2015}), $\theta$ is the complex phase of $Q_e$, and $f_r$ is the resonance frequency. Figure \ref{fig:fitting} illustrates a typical example of low power resonator response fitting.

\section{Results and Discussion}
The measured transmission data $S_{21}(f)$ obtained from the VNA were processed using a circle-fit routine to extract the resonance frequency $f_r$, loaded quality factor $Q_l$, and coupling quality factor $Q_e$. The internal quality factor $Q_i$ was then calculated via Eq.~\ref{eq:Q_load}. Each dataset corresponded to a different probe power level, enabling analysis of power-dependent losses.

The microwave power reaching the resonator chip differs from the VNA output power due to attenuation and losses along the input line. These include fixed cryogenic attenuators, frequency-dependent coaxial cable losses, and insertion losses from intermediate components such as filters, DC blocks, and connectors, all of which contribute to the total line attenuation. The power delivered to the device plane reads 
\begin{equation}
P_{\mathrm{chip},j}\,[\mathrm{dBm}] = P_{\mathrm{VNA},j}\,[\mathrm{dBm}] - L_{\mathrm{tot}}\,[\mathrm{dB}]
\label{eq:power}
\end{equation}
where $L_{\mathrm{tot}} = L_{\mathrm{atten}} + L_{\mathrm{cable}} + L_{\mathrm{insertion}}$ represents the total attenuation (Table I shows each contribution to the total attenuation value in the used setup).

To compare resonators with different frequencies and coupling rates on equal footing, the internal loss is analyzed as a function of the average photon number $\bar{n}$ stored in the mode, namely,
\begin{equation}
\bar{n}= \frac{2 P_{\mathrm{chip},j}[{\rm W}] Q_l^2}{\hbar \omega_r^2 Q_e},
\label{eq:n_bar}
\end{equation}
rather than the incident power at the (VNA) output. 
Expressing the data in terms of $\bar{n}$ improves comparison across resonators by converting the applied drive power into an estimate of intracavity photon number, which depends on the coupling and line attenuation calibration. In this way, $Q_i$ is still extracted from each resonance lineshape~\cite{Megrant_2012, Place_2021,McRae_2020,Gao_2008_Thesis}. It is emphasized that uncertainties in $L_{\mathrm{tot}}$ do not affect the extracted %
$Q_i$ from each lineshape, but do affect the absolute calibration of $\bar{n}$. 

\begin{table*}[t]
    \centering
    \label{tab}
    \begin{threeparttable}
        \caption{Input Line Attenuation}
        \vspace{0.2cm} 
         \begin{tabular*}{\textwidth}{@{\extracolsep{\fill}}l c l@{}}
            \hline \hline 
            \textbf{Component} & \textbf{Atten. (dB)} & \textbf{Notes} \\ 
            \hline %
            
            Direct Attenuators & 62.0 & Fixed value (Sum of discrete cryogenic attenuators). \\
             \\[0.5ex] 
             
            RLC F-30-8000-R LPF & 0.35 & \textbf{Spec:} 0.35 dB Max\tnote{a} \\
             \\[0.5ex]
             
            Eccosorb IR Filter & 2 & \textbf{Spec:} $<1.8$ dB at 10 GHz (4K)\tnote{b} \\
             \\[0.5ex]
             
            Room-Temp Coax & 3 & 3-4m cables plus connectors. \\
             \\[0.5ex]
             
            Bluefors Internal Coax & 2 & \! \\
            \\[0.5ex]
            
            \textbf{Total} & \textbf{69.35} & \textbf{Used: $\sim$69 dB} \\
            
            \hline \hline 
         \end{tabular*}
        
        \begin{tablenotes}
            \footnotesize
            \item[a] RLC Electronics, ``F-10/F-30 Series Low Pass Filters Specifications''. Datasheet Table, Row ``F-30-8000''.
            \item[b] Quantum Microwave, ``QMC-CRYOIRF-002 Cryogenic IR Filter''. Spec: Insertion loss $<1.8$ dB at 10 GHz @ 4K.
        \end{tablenotes}
    \end{threeparttable}
\end{table*}

\begin{figure}[h]
\centering
\includegraphics[width=\columnwidth]{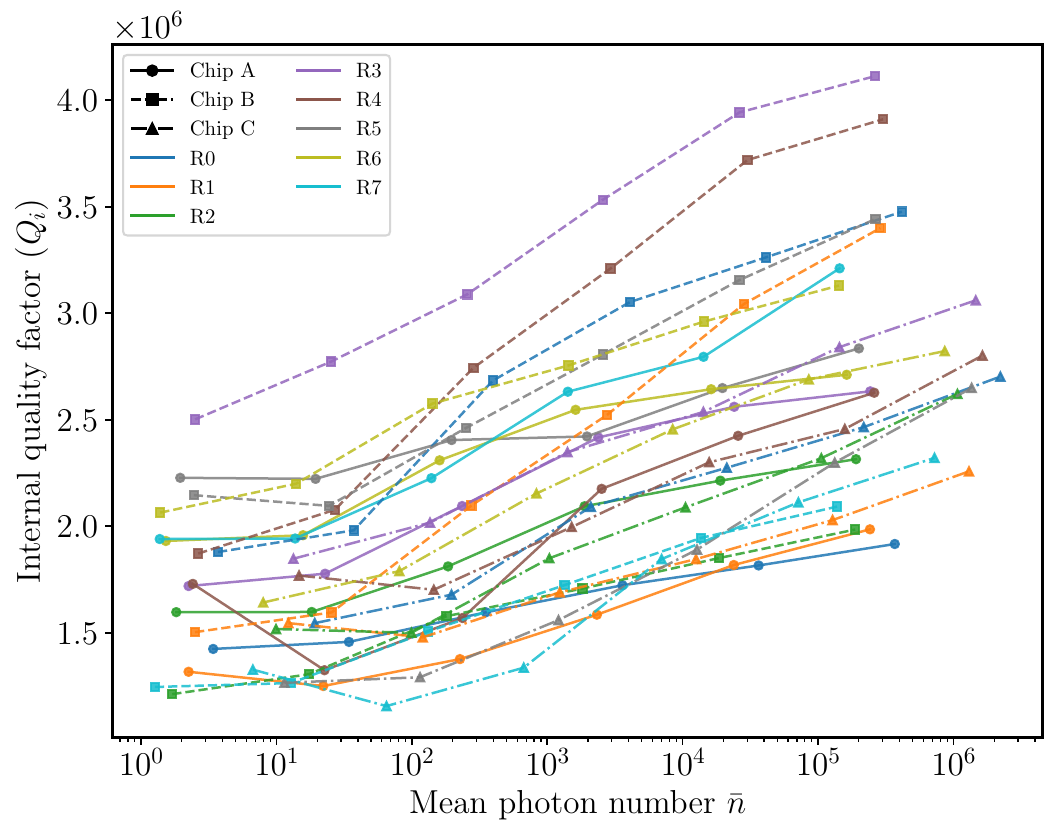}
\caption{Internal quality factor $Q_i$, of three chips fabricated and tested separately, as a function of mean photon number $\bar{n}$ for all resonators. The increase of $Q_i$ with $\bar{n}$ reflects saturation of power-dependent loss.}
\label{fig:n_vs_Q}
\end{figure}

To obtain a clear resonator‑independent view of dissipation, the internal quality factor $Q_i$ as a function of $\bar{n}$ for resonators on three chips fabricated and tested separately (different cryogenic cooldown) is shown in (Fig.~\ref{fig:n_vs_Q}). In this representation, all resonators on the chips exhibit a characteristic feature: in the single‑photon regime $\bar{n} \approx 1$, the internal quality factor is reduced and only weakly dependent on drive power, while $Q_i$ increases smoothly as $\bar{n}$ is increased. Such behavior is consistent with saturation of an ensemble of TLS defects that are expected to dominate loss at low photon number in many superconducting resonators.

The dependence of $Q_i$ on $\bar{n}$ is well described by the standard resonant TLS model~\cite{Gao_2008_Thesis,Gao_2008,McRae_2020}. Because independent loss channels add linearly in inverse-$Q$ space, the effective internal loss can be written as
\begin{subequations}
\label{eq:tls_model}
\begin{equation}
    \frac{1}{Q_i(\bar{n})} 
    = 
    \frac{1}{Q_0} + \frac{1}{Q_{\mathrm{TLS}}}\frac{1}{\sqrt{1+\bar{n}/n_c}}, 
    \label{eq:tls_model_Q}
\end{equation}
where the following effective mode‑averaged loss tangents are defined for the power‑independent background channel, the saturable TLS channel, and the aggregate internal dissipation, respectively:
\begin{align}
    \tan\delta_{0,\mathrm{eff}} &= \frac{1}{Q_0}, \label{eq:tandelta_a} \\
    \tan\delta_{\mathrm{TLS},0,\mathrm{eff}} &= \frac{1}{Q_{\mathrm{TLS}}}, \label{eq:tandelta_b} \\
    \tan\delta_{\mathrm{eff}}(\bar{n}) &\equiv \frac{1}{Q_i(\bar{n})}. \label{eq:tandelta_c}
\end{align}
In this notation, the TLS model of Eq.~\eqref{eq:tls_model_Q} is expressed as
\begin{equation}
    \tan\delta_{\mathrm{eff}}(\bar{n}) 
    = 
    \tan\delta_{0,\mathrm{eff}} + \frac{\tan\delta_{\mathrm{TLS},0,\mathrm{eff}}}{\sqrt{1+\bar{n}/n_c}}.
    \label{eq:tls_model_tan}
\end{equation}
\end{subequations}

Here $\tan\delta_{\mathrm{eff}}(\bar{n})$ represents an effective mode loss aggregating all internal dissipation rather than an intrinsic material loss tangent, and thus should not be directly compared to microscopic loss tangents without participation‑ratio analysis. In the expressions above, $Q_0$ denotes the power-independent background quality factor, $Q_{\mathrm{TLS}}$ sets the low-power TLS loss scale, and $n_c$ is the characteristic photon number for TLS saturation. 

The square-root dependence arises from the nonlinear steady-state solution of driven TLS dipoles obtained from the Bloch equations~\cite{McRae_2020,Faoro_2012}, which yields a TLS loss tangent of the form $\tan \delta_{\mathrm{TLS}}(E) \propto 1 / \sqrt{1 + E^2/E_c^2}$, with $E$ the local electric field and $E_c$ the critical electric field at which the TLS polarization saturates; mapping $E_c^2$ to stored energy or photon number produces the $1/\sqrt{1 + \bar{n}/n_c}$ dependence used here. The quantity $n_c$ therefore encodes an effective combination of TLS dipole moments, relaxation times, and the spatial distribution of the resonator's electric field~\cite{McRae_2020,Gao_2008_Thesis, Lindstrom_2009}.

This TLS model predicts two limiting regimes that match the experimental trends. For $\bar{n} \ll n_c$, the TLS ensemble is unsaturated, the TLS term $1/Q_{\mathrm{TLS}}$ dominates, and the total internal loss is approximately $1/Q_i \approx 1/Q_0 + 1/Q_{\mathrm{TLS}}$. For $\bar{n} \gg n_c$, the TLS contribution is suppressed as $1/\sqrt{\bar{n}/n_c}$, and the internal loss asymptotically approaches the background level $1/Q_i \to 1/Q_0$, yielding a high-power loss floor set by non-TLS mechanisms.

\begin{figure}[!h]
\includegraphics[width=\columnwidth]{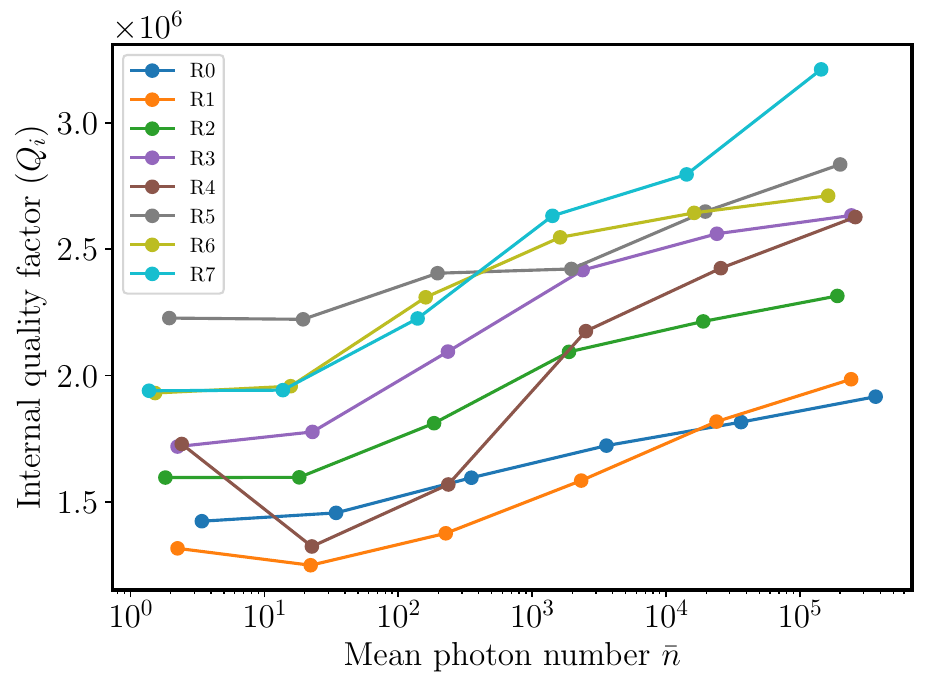}
\caption{Internal quality factor $Q_i$ as a function of mean photon number $\bar{n}$ for all resonators on Chip A. The increase of $Q_i$ with $\bar{n}$ reflects saturation of power-dependent loss.}
\label{fig:n_vs_Q_chipA}
\end{figure}
\begin{figure}[!h]
\includegraphics[width=\columnwidth]{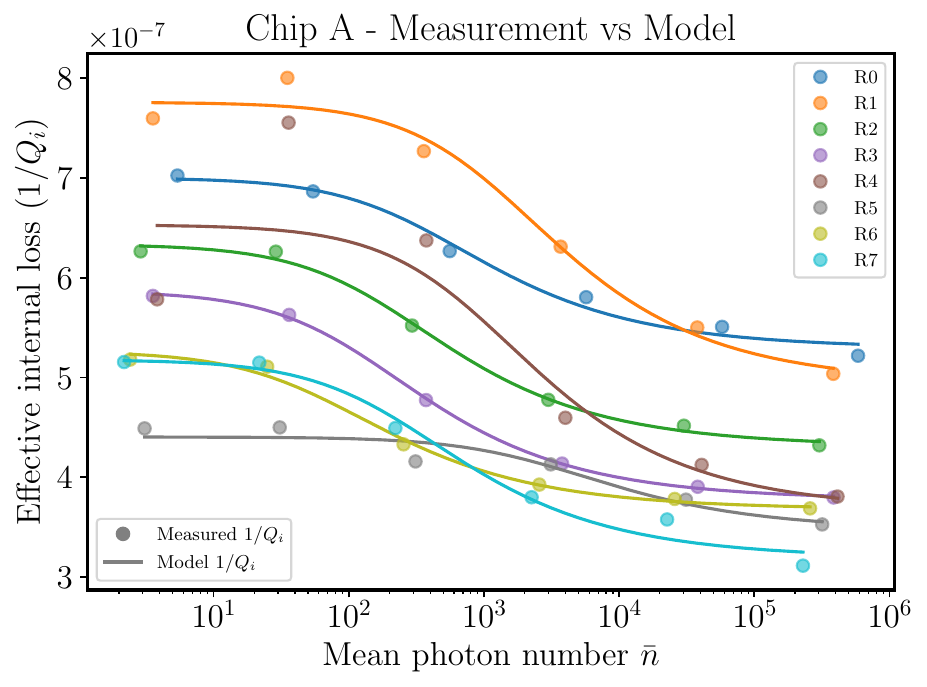}
\caption{Effective internal loss $1/Q_i$ of Chip A as a function of mean photon number $\bar{n}$.}
\label{fig:decomp_chip_a}
\end{figure}
\begin{figure}[!h]
\includegraphics[width=\columnwidth]{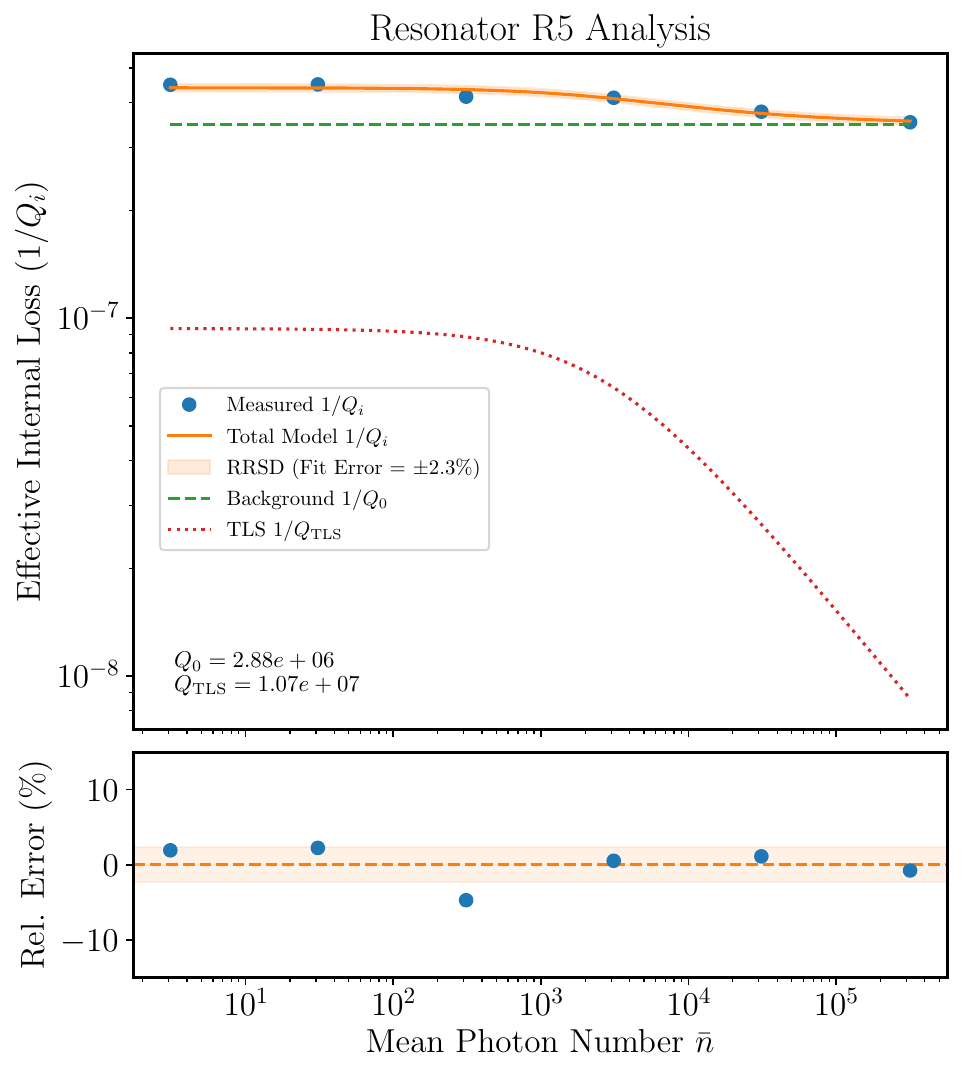}
\caption{Effective internal loss $1/Q_i$ of resonator R5 of Chip A as a function of mean photon number $\bar{n}$. Top: The total fitted loss (solid line) is decomposed into a constant background term $1/Q_0$ (dashed) and a saturable Two-Level System (TLS) contribution (dotted) governed by the Standard Tunneling Model. Bottom: Relative residuals (percentage error) of the measured data from the TLS model.}
\label{fig:decomp_r5}
\end{figure}
\begin{figure*}[t]
\includegraphics[width=\textwidth]{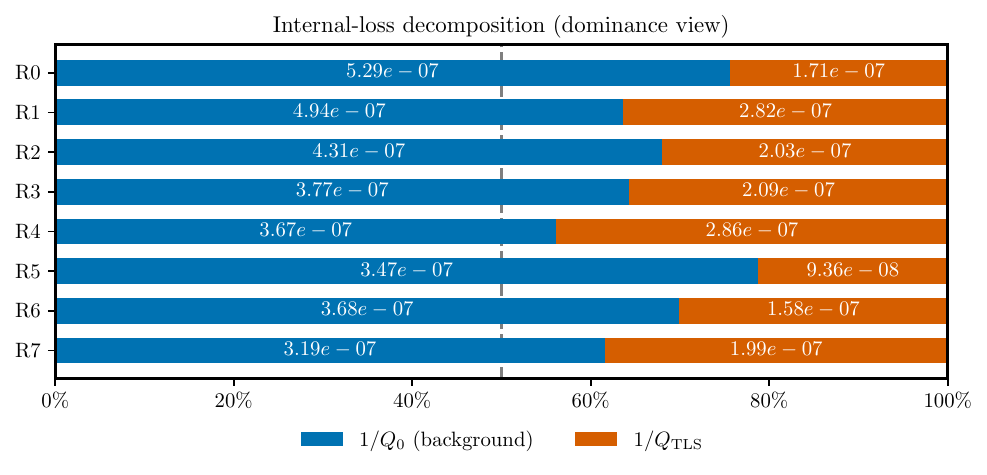}
\caption{Internal loss decomposition of resonators R0–R7 of Chip A. The effective internal loss $1/Q_i$ is expressed as the sum of a background contribution $1/Q_0$ and a TLS-related contribution $1/Q_{\mathrm{TLS}}$. The bars show the normalized contributions of each loss channel, with in-bar labels indicating the corresponding absolute loss values.}
\label{fig:loss_decomp_all}
\end{figure*}

The main results presented here (Figs. \ref{fig:n_vs_Q_chipA}, \ref{fig:decomp_chip_a}, \ref{fig:decomp_r5}, and \ref{fig:loss_decomp_all}) are obtained from eight coplanar waveguide resonators fabricated on the same chip (named chip A), spanning frequencies from approximately 5.0 GHz to 6.4 GHz with a spacing of about 200 MHz. All resonators share an identical substrate, metal stack, lithographic patterning, etching, and cleaning process, and were measured in the same package and cryogenic cooldown. This common experimental context allows variations in internal quality factor to be interpreted primarily in terms of resonator frequency, coupling design, and intrinsic loss mechanisms, rather than process-to-process or cooldown-to-cooldown variability.

To visualize the different contributions, we selected the chip A, as its data (highlighted in Fig.~\ref{fig:n_vs_Q_chipA}) are representative of the typical values obtained across the three measured chips. As an initial validation, Figure~\ref{fig:decomp_chip_a} illustrates the aggregate agreement between the measured internal losses ($1/Q_i$) and the fitted total loss model across all functional resonators on this chip. Figure ~\ref{fig:decomp_r5} shows the loss decomposition for a representative resonator, plotted as $1/Q_i(\bar{n})$ together with the fitted constant background $1/Q_0$ and the TLS term. In this inverse-Q representation, the TLS contribution is dominant at low $\bar{n}$, while at high $\bar{n}$ it collapses and the total loss converges to the constant background, as expected from the TLS saturation model. Systematic, photon-number-dependent structure in the residuals beyond this simple two-component form would indicate additional mechanisms such as power-dependent quasiparticle generation, heating, non-TLS nonlinearities (e.g. Kerr effects), multiple TLS populations, or artifacts of the measurement chain. For this resonator, the fitted parameters yield $Q_0 \approx 2.88\times10^6$, $Q_{\mathrm{TLS}} \approx 1.07\times10^7$, and a critical photon number $n_c \approx 1.72\times10^3$, values that are physically reasonable~\cite{Poorgholam-Khanjari_2025} and consistent with the trends observed across the chip.

The quality of the TLS-based description is further assessed by examining the fit residuals for the same resonator, shown in (Fig.~\ref{fig:decomp_r5}). To quantify the deviation of the measured internal loss from the standard Two-Level System (TLS) model, we calculate the Relative Residual Standard Deviation (RRSD). For each resonator, the relative error is computed at each measured mean photon number ($\bar{n}$) as the difference between the measured loss ($1/Q_{i,\text{meas}}$) and the model fit ($1/Q_{i,\text{model}}$), normalized by the measured loss. The RRSD is defined as the standard deviation of these percentage-based residuals. Unlike an absolute root-mean-square error, which is disproportionately weighted by the high-loss (low-power) regime, the RRSD provides a dynamically scaled, power-independent uncertainty metric. This yields a single relative fit error ($\pm X\%$) that represents the scatter of the experimental data around the idealized model. The residuals remain small over the full photon-number range and exhibit no systematic dependence on $\bar{n}$, indicating that no obvious additional power-dependent loss channel is required to describe the data within the experimental resolution. This supports the use of the same TLS framework to parameterize the photon-number-dependent loss behavior of all resonators on the chip.

The internal loss decomposition of the eight resonators on chip A (Fig.~\ref{fig:loss_decomp_all}) shows that, for all %
resonators, the total dissipation is dominated by the background channel $1/Q_0$, with the TLS-related contribution $1/Q_{\mathrm{TLS}}$ remaining secondary but non-negligible. Quantitatively, $1/Q_0$ accounts for roughly 60--80\% of the total internal loss, while $1/Q_{\mathrm{TLS}}$ contributes the remaining 20--40\%, indicating that none of the resonators is purely TLS-limited under the present measurement conditions. Although this overall hierarchy between background and TLS loss is consistent across the chip, the relative fractions vary from resonator to resonator: for example, R5 ~\ref{fig:decomp_r5} exhibits a comparatively small TLS contribution, with $1/Q_{\mathrm{TLS}}$ almost an order of magnitude lower than $1/Q_0$, whereas R1 and R7 show larger TLS fractions approaching $\sim 40\%$ of the total loss. These variations occur despite nominally identical fabrication and measurement conditions, suggesting sensitivity to factors such as resonant frequency, local electric-field distribution, and the participation of lossy interfaces. Collectively, the data indicate that TLS loss has been partially mitigated but remains relevant for the fundamental mode, while a power-independent background establishes a common floor for the achievable internal quality factor, consistent with trends reported for high-$Q$ planar CPW resonators where improvements in interface quality progressively reveal non-TLS loss mechanisms as the dominant limitation. While $n_c$ and $Q_{\mathrm{TLS}}$ provide insight into the effective dipole moments and surface participation, the background loss $Q_0$ requires careful interpretation regarding the electromagnetic environment.

\begin{figure}[!h]
\includegraphics[width=\columnwidth]{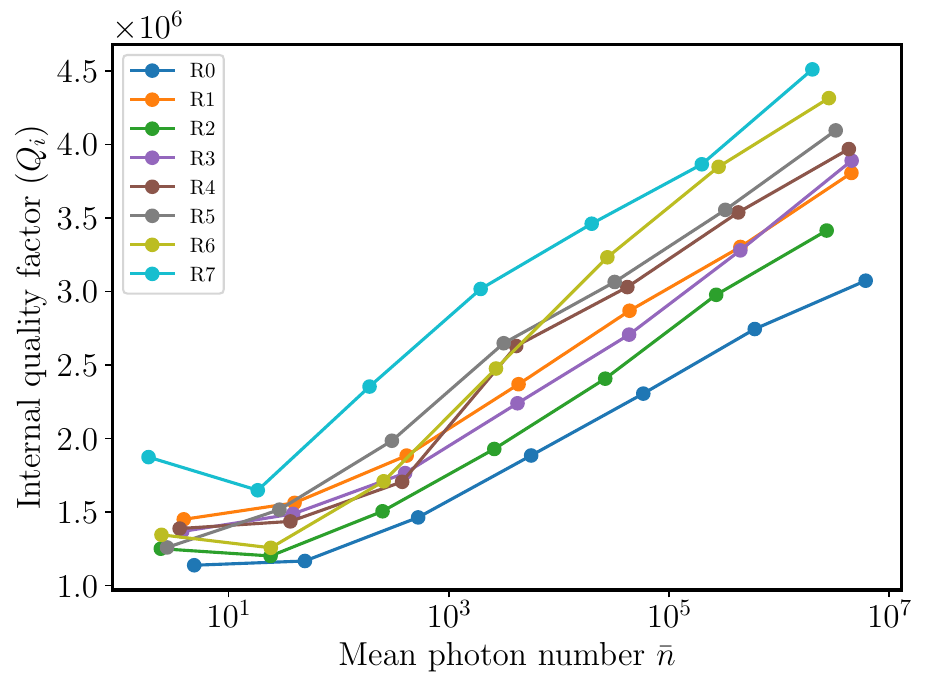}
\caption{Internal quality factor $Q_i$ as a function of mean photon number $\bar{n}$ for all resonators on standard Nb chip.}
\label{fig:nb_qi}
\end{figure}

Because the resonators on Chip A share the same material stack, fabrication flow, packaging, and cooldown history, the observed variation in the extracted high-power background quality factor $Q_0$  (Fig. \ref{fig:nb_qi}) is more likely to reflect geometry, and environment-dependent loss mechanisms than fabrication-induced material differences. In particular, stronger coupling structures can enhance unintended interaction with the surrounding electromagnetic environment, including radiation into package modes, excitation of slotline modes, and coupling to ground-plane modes, all of which are recognized dissipation channels in superconducting microwave circuits and packaging studies. More generally, the resonator-to-resonator variation in $Q_0$  is consistent with a background loss that is influenced by the device electromagnetic environment rather than by the material stack alone, since resonators at different frequencies can couple differently to parasitic enclosure modes and other frequency-selective loss channels~\cite{Megrant_2012,Bruno_2015,McRae_2020,Lienhard_2019,Reagor_2013,Zmuidzinas_2012,Huang_2021}.
Therefore, while $Q_{\mathrm{TLS}}$ serves as the primary metric for evaluating the quality of the Nb/Ta fabrication and surface passivation, $Q_0$ is expected to be more strongly influenced by conductor loss, radiation, and packaging-related dissipation.

\begin{figure}[!h]
\includegraphics[width=\columnwidth]{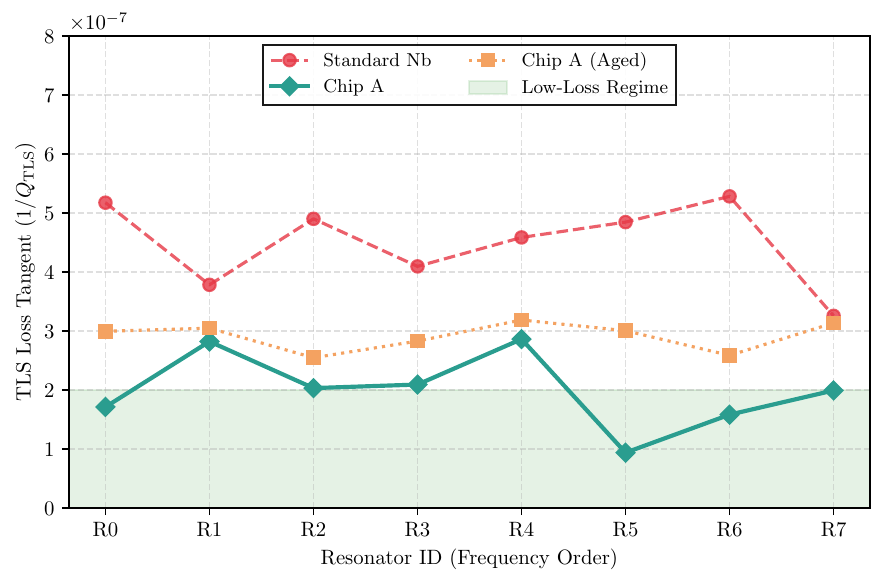}
\caption{Extracted TLS loss tangent ($1/Q_{\mathrm{TLS}}$) for eight resonators (R0--R7) comparing a standard Nb baseline, fresh Chip A (Nb/Ta), and the same chip after six months of aging/storage.}
\label{fig:tls_comp}
\end{figure}

To further contextualize the performance and robustness of the Nb/Ta material system, we fabricated a reference Nb-only resonator chip using an identical process flow (excluding the Ta capping step) and re-evaluated Chip A resonators after six months of storage. These measurements were performed in a different dilution refrigerator with a base temperature of 23 mK (vs. 10 mK for the primary chips) and input line attenuation of 60 dB (vs. 62 dB). Figure~\ref{fig:nb_qi} shows the internal quality factor $Q_i$ as a function of $\bar{n}$ for the Nb-only chip, while Figure~\ref{fig:tls_comp} compares the extracted TLS loss ($1/Q_{\mathrm{TLS}}$) for the fresh Nb/Ta devices, aged Nb/Ta devices, and fresh Nb-only devices. The fresh Nb/Ta resonators exhibit the lowest TLS dissipation, with typical $Q_{\mathrm{TLS}}$ values in the range $4 \times 10^6$–$1 \times 10^7$, while the Nb-only resonators show significantly higher TLS loss, with a mean $Q_{\mathrm{TLS}} \approx 2.3 \times 10^6$. After six months, the Nb/Ta devices display an increase in $1/Q_{\mathrm{TLS}}$, indicating gradual degradation of the Ta-based passivation layer, yet their TLS loss remains lower than that of the freshly fabricated Nb reference, with aged Nb/Ta resonators exhibiting a mean $Q_{\mathrm{TLS}} \approx 3.5 \times 10^6$. Taken together, these measurements support the conclusion that Ta encapsulation improves the initial resonator loss budget and provides a meaningful degree of robustness against long‑term aging effects.

\section{Conclusion}
Coplanar waveguide resonators based on magnetron-sputtered Nb with a Ta cap on high-resistivity Si were characterized in a single cooldown using power‑dependent spectroscopy on eight resonators patterned on the same chip. The internal quality factor reaches the few‑million range near the single‑photon regime and exhibits the expected increase with drive power, which is well captured by a standard TLS saturation model that decomposes the internal loss into a power‑independent background term and a saturable TLS contribution.

The fitted loss decomposition shows that on the Nb/Ta chip the background channel $1/Q_0$ dominates over the TLS‑related contribution $1/Q_{\mathrm{TLS}}$, indicating that these resonators are not strongly TLS‑limited under the present conditions and that geometry‑ and environment‑dependent mechanisms are an important factor in setting the residual loss floor. The relatively small TLS contribution is consistent with effective suppression of interface‑related TLS loss by the Ta capping layer, while background loss remains the primary target for further improvement.

To assess the robustness of this gain, the Nb/Ta devices are compared to a reference Nb device fabricated with an identical process flow and to the same Nb/Ta chip after six months of storage. The Nb‑only resonators exhibit a lower $Q_{\mathrm{TLS}}$ than the fresh Nb/Ta devices, consistent with stronger TLS‑related loss at the uncapped Nb surface. The aged Nb/Ta devices show a moderate increase in $1/Q_{\mathrm{TLS}}$, yet their TLS loss remains below that of the freshly fabricated Nb reference. This suggests that Ta encapsulation not only improves the initial loss budget but also provides a degree of robustness against long‑term ageing effects.

Although comprehensive interface studies, such as microscopic analysis, X-ray photoelectron spectroscopy, and systematic Ta-thickness optimization, remain for future work, the present results support Ta encapsulation as a viable route to suppressing TLS-related losses in planar superconducting circuits. Within this framework, further gains in coherence are expected to depend on concurrent optimization of the electromagnetic environment and device layout, together with continued refinement of materials and interfaces, in order to fully exploit the potential of Nb/Ta-based platforms for scalable superconducting quantum hardware.

\section{Acknowledgments}
The authors acknowledge Gianluigi Catelani for a critical review. The authors thank Tatiana Kazieva, Laura Gallego, and Varun Madhavan for their dedicated support in laboratory organization, experimental setup, and essential background activities that enabled the smooth execution of this work. The authors further acknowledge Frederico Brito for supervising the project, contributing to its conceptual development, and providing critical feedback on the manuscript. During manuscript preparation, large language models (LLMs) were used solely to improve readability and linguistic clarity. These tools were not used for scientific analysis, data generation, conceptual development, or interpretation of results. All scientific judgments and statements presented in this work originate from the authors.

\section{Author Contributions}
AA conceptualized the study, process development and devices fabrication, conducted the experimental measurements and data analysis, and prepared the original manuscript draft. JV performed design and simulations of devices components. FR contributed to device fabrication. AZ assisted with the experimental setup and measurements. All authors have read and contributed to the final version of the manuscript. %

\section{References}
\bibliographystyle{apsrev4-2}
\bibliography{Sections/ref_v1}

\end{document}